\documentstyle[prl,twocolumn,aps,graphicx]{revtex}
\begin{document}
\newcommand {\be}{\begin{equation}}
\newcommand {\ee}{\end{equation}}
\newcommand {\bea}{\begin{eqnarray}}
\newcommand {\eea}{\end{eqnarray}}
\newcommand {\nn}{\nonumber}

\draft
\twocolumn[\hsize\textwidth\columnwidth\hsize\csname 
@twocolumnfalse\endcsname
%
%
%

\title{Instabilities at $[110]$ Surfaces of $d_{x^2-y^2}$ Superconductors }

\author{Carsten Honerkamp$^{(1)}$, Katsunori Wakabayashi$^{(2)}$, and
Manfred Sigrist$^{(2)}$}

\address{$^{(1)}$ Theoretische Physik,
ETH-H\"onggerberg, CH-8093 Z\"urich, Switzerland,\\
$^{(2)}$ Yukawa Institute for Theoretical Physics, Kyoto University,
Kyoto 606-8502, Japan}

\date{\today}
\maketitle

\begin{abstract} 
We compare different scenarios for the low temperature 
splitting of the zero-energy peak in the local density of states at
(110) surfaces of $d_{x^2-y^2}$-wave superconductors, observed by 
Covington  
et al. (Phys.Rev.Lett. {\bf 79} (1997), 277). Using a tight binding model in the
Bogolyubov-de Gennes treatment	 we find a surface phase transition towards a 
time-reversal symmetry breaking surface state carrying spontaneous currents and
an $s+id$-wave state. Alternatively, we 
show that electron correlation 
leads to a surface phase transition towards a magnetic state
corresponding to a local spin density wave state.  
\end{abstract}

\vskip1pc]
\narrowtext
A large number of experiments have established that the Cooper pair
wavefunction has $ d_{x^2-y^2} $-wave symmetry in high-temperature
superconductors. Decisive information came from 
probes which are sensitive to the internal phase structure of the pair 
wavefunction. Besides experiments based on the
Josephson effect \cite{josephson} 
also the surface Andreev bound states (ABS)\cite{hu}  observable in
quasiparticle tunneling can be counted among the strongest 
experimental tests of this type. For a superconductor with
pure $d_{x^2-y^2}$ symmetry these ABS should be most pronounced at
[110] surfaces and lie  exactly at the Fermi energy (zero energy). They lead to a
rather sharp zero-bias anomaly in the I-V tunneling characteristics
which reflects the surface quasiparticle density of states (DOS) 
\cite{greene}. In addition low temperature anomalies in the
penetration depth have been interpreted as evidence for the existence
of the zero-energy ABS\cite{walter}.  
An interesting twist in the view of the ABS occurred when Covington
et al.\cite{covington}  observed the spontaneous split of the single
zero-energy peak (ZEP) into two peaks at finite voltage equivalent to an energy of approximately 
10$\%$ of the superconducting gap below $ T = 7 K $. 
Fogelstr\"om et al.\cite{fogelstrom}
interpreted this in terms of a spontaneous violation of time-reversal
symmetry breaking (TRSB) by the admixture of a sub-dominant $s$-wave component
close to the surface. The split ABS is the result of the opening
of a small gap in the quasiparticle spectrum at the
surface and can be interpreted as a Fermi surface (FS)
instability.\cite{trsbreview,yip}   
TRSB is not the only way to shift the ZEP to finite
energies. Many local FS instabilities could yield
the same effect and the one with the largest energy gain, i.e. the
highest critical temperature, would finally govern the surface state.
In this letter we discuss the instability due to correlation effects
among the quasiparticles. The zero-energy ABS consists of degenerate
states with a charge current running 
parallel to the surface (in-plane) in both directions (the {\em directional degeneracy}) and with both
spin up and down (the {\em spin degeneracy}). The TRSB state lifts the
directional degeneracy of charge currents by 
admixing a subdominant s-wave component to the d-wave pairing state and a
spontaneous finite current appears.\cite{trsbreview} On the other hand, the spin
degeneracy can be lifted yielding a spin density wave-like state at
the surface, although magnetic ordering is
absent in the bulk.\cite{waka} This instability is driven by the repulsive
electron-electron interaction responsible for the strong
antiferromagnetic spin fluctuations in the underdoped region of
high-temperature superconductors. From this point of view the magnetic
instability represents an equally probable way to lift the
degeneracy of the zero-energy states. 

\begin{figure} \begin{center} 
\includegraphics[width=.5\textwidth]{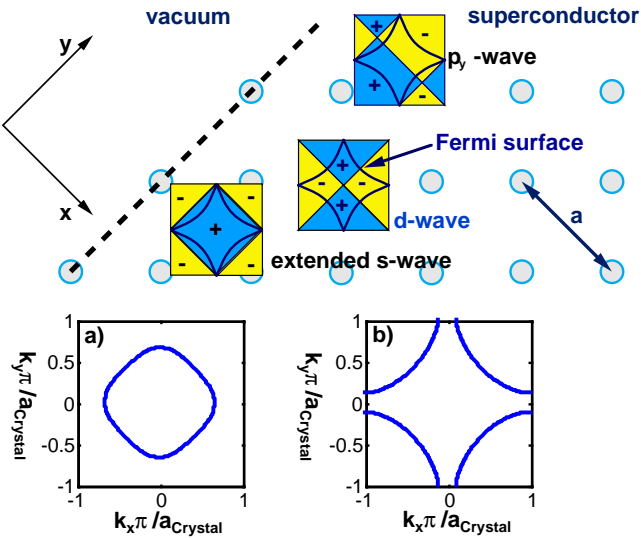}
\end{center} \caption{Geometry and coordinate system of the [110] 
boundary.  The symmetries of the relevant gap functions are 
symbolized by the $+$ and $-$ regions in the Brillouin zones. a) Fermi surface
for the {\em regular FS} parameters given in the text. b) Fermi surface
for {\em singular FS}  parameters. In our notation $a=\sqrt{2} a_{Crystal}$}
\label{geom}
\end{figure}

In the following we analyze the properties of [110]-oriented surface
of the $ d_{x^2-y^2}$-wave superconductor 
on a square lattice forming a strip of finite width (infinitely long along
[1,-1,0]-direction) such that we have two surfaces.
The translationally invariant
direction is denoted by the $ y$-axis and the surface normal
direction by the $x$-axis (Fig.1). We describe this system by a tight-binding
model with nearest ($ t $)  and next-nearest ($t'$) neighbor
hopping, and include an onsite repulsive and a spin-dependent nearest
neighbor interaction. The latter generates the superconducting state
while the former introduces rather magnetic correlations. The corresponding
Hamiltonian has the form

\begin{equation} \begin{array}{ll}
{\cal H} = & -t \sum_{\langle {\bf x}, {\bf x}' \rangle,s}
c^{\dag}_{{\bf x}s} c_{{\bf x}'s} - t' \sum_{\langle {\bf x}, {\bf x}'
\rangle',s} c^{\dag}_{{\bf x}s} c_{{\bf x}'s} \\ & \\
& + J \sum_{\langle {\bf x}, {\bf x}' \rangle} {\bf S}_{{\bf x}} \cdot
{\bf S}_{{\bf x}'} + U \sum_{{\bf x}} n_{{\bf x} \uparrow} n_{{\bf x}
\downarrow}. 
\end{array} \end{equation}
The interaction terms are decoupled by meanfields involving the onsite
charge $ n(\bf x) = \sum_s \langle c^{\dag}_{{\bf x}s} c_{{\bf x}s}
\rangle $, the onsite magnetic moment $ m (\bf x) = \sum_s s \langle
c^{\dag}_{{\bf x}s} c_{{\bf x}s} \rangle $ and the pairing on nearest
neighbor sites $ \Delta_{s,s'} ({\bf x},{\bf x}') = \langle c_{{\bf
x}s} c_{{\bf x}'s'} \rangle $ which include both 
$S = 0$ and $S =  1$ pairing. However, spin-triplet pairing
is suppressed due the repulsive nature of 
the interaction in that channel, if we chose $J>0$. It is sufficient for our
analysis to include pairing meanfields $ \Delta_{s,-s}({\bf x},{\bf
x}')$ on the bonds. 
This yields basically four types of gap functions, two with singlet
($s$- and $d$-wave) and two
with spin-triplet ($p_x$- and $p_y$-wave with $ S_z =0 $)
pairing. Among these we find the 
$d_{x^2-y^2}$-wave (or $d_{xy}$ in our rotated coordinates) state as
dominant bulk phase.    
An on-site $s$-wave without nodes is excluded here, instead we obtain
an extended $s$-wave state. 
In Fig.\ref{geom} we 
symbolize the corresponding gap functions of these three relevant 
pairing states in momentum space.
All the position dependent mean-fields are then determined
selfconsistently together with current densities and the vector potential solving
the corresponding Bogolyubov-deGennes equations. For the solution of
the Maxwell equation we impose the 
boundary conditions of zero magnetic field at the surface and vanishing vector
potential in the bulk.
We consider two typical cases for the band structure: (1) band filling
approximately $ 60 \% $ and  $ t' = 0 $, where the FS has nearly
circular shape ({\it regular FS}); (2) band filling nearly $ 85 \% $
and $ t' = - 0.3 t $ with the FS close to van Hove singularities (VHS) 
at $ (\pi,0) $ and $ (0, \pi ) $ ({\it singular FS}). The latter case
represents the situation in underdoped, the former rather in overdoped 
region of the cuprate phase diagram.

\begin{figure} \begin{center} 
\includegraphics[width=.48\textwidth]{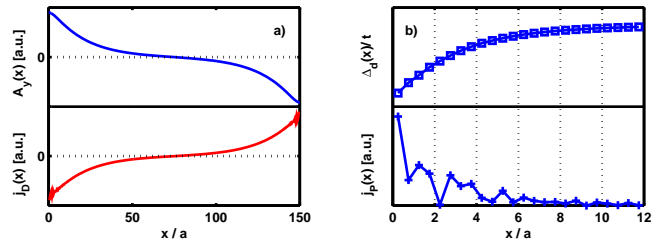}
\end{center} \caption{Spontaneous surface current state: a): Vector potential $A_y$ and
diamagnetic Meissner current density $j_D$. b): $d$-wave gap function
$\Delta_d$ and the paramagnetic surface current density $j_P$
carried by the Andreev bound states (system width $150a$, $\mu =-t$, $t'=0$,  $\lambda \approx 18 a$ and $T=0.001t$ yielding critical temperatures
$T_c^{SSC}= 0.006 t$ and $T_c^d = 0.11t $).} 
\label{spocu}
\end{figure}

First we consider the {\em regular FS}. The instability in
the charge channel can be examined by setting $U=0$. With the
pairing symmetry restricted to the $d_{x^2-y^2}$-wave 
channel, the only instability possible is by generating a spontaneous 
current running along the surface.
The mechanism can be understood in the following way. Assume a
vector potential $A_y(x)$ along the surface. This will shift the ABS carrying 
charge in one direction  to negative energies, and, thus, creating a
paramagnetic surface current  
which decays into the bulk  on the distance of the superconducting coherence
length $\xi$. The energy 
gain $-\int dx j_y(x)A_y(x)$ through  the surface currents is basically
$\propto A_y$, while the energy costs from the required Meissner screening currents are only
$\propto A_y^2$. This then leads to a minimum of the total energy for a
certain finite $A_y(x)$. We will call this TRSB state the spontaneous
surface current (SSC) state. The spatial dependence of  the vector potential 
$A_y(x)$, diamagnetic  and paramagnetic current densities and the $d$-wave gap
function are shown in Fig.\ref{spocu}.  From quasi-classical
calculations\cite{honunpub} one expects a critical temperature
$T_c^{SSC}$ of the order $(\xi / \lambda) T_c^d$, i.e. rather small for a typical high-$T_c$
superconductor. Our numerical results show $T_c^{SSC} = 0.006t $ or
$T_c^{SSC} / T_c^d \approx \frac{1}{2}\xi / \lambda$. 
 This low transition
temperature indicates that the split of the ABS levels
is rather small and barely visible in the surface DOS. We conclude
that this surface instability is a poor candidate in order to explain the experimental observations.

\begin{figure}[t] \begin{center} 
\includegraphics[width=.5\textwidth]{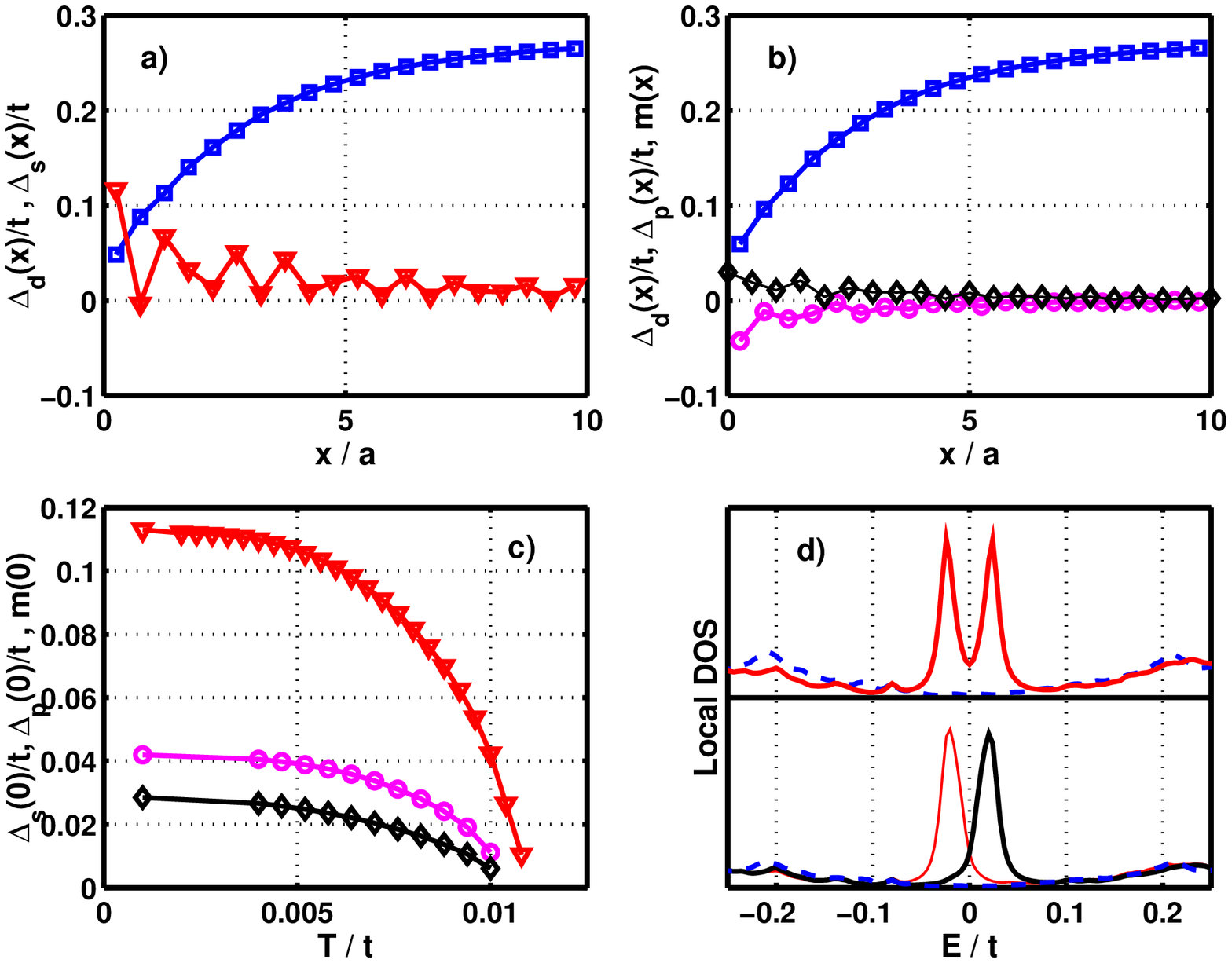}
\end{center} \caption{Regular FS scenario (i.e. $t'=0$, $J=2 t$, $\mu 
=-t$ and $T_c^d= 0.11t$): a) $d$-wave gap
(squares) and TRSB $s$-wave 
admixture (triangles) at
$T=0.001 t$ for $U=0$. b) $d$-wave gap, magnetization
(diamonds) and TRSB $p$-wave admixture (circles) at
$T=0.001 t$ for $U=t$. c): Temperature dependence of the $s$-wave gap
function, $p$-wave 
gap function  and 
magnetization on the first site. d) upper panel: Local DOS in
the $s+id$ state at the surface (solid line) and in the bulk (dashed
line); lower panel: Local DOS in
the magnetic surface state at the surface (solid lines, spin-up and spin-down
DOS separate) and in the bulk (dashed line).}
\label{all_tp0}
\end{figure}

Now we include a finite $s$-wave component with a relative phase
of $\pm \pi/2$
with respect to the $d$-wave gap ($s \pm i d $). This phase is chosen to maximize
the condensation energy of the surface state. Since this TRSB $s$-wave admixture
also lifts the charge degeneracy by changing the Andreev reflection
properties for the states with left and right going charge currents, it
leads naturally to a net surface
current. In the presence of a finite attractive $s$-wave coupling this
will lead to an additional contribution in the gap function 
increasing the energy gain and resulting in a wider splitting of the
ZEP. As a consequence the critical
temperature for this $s+id$ state is higher than for the SSC
state without $s$-component. 
Spatial and temperature dependence of
$s$-admixture and vector potential are shown in Fig.\ref{all_tp0}. For our
choice of parameters, the split in the surface DOS has similar size
relative to the bulk gap value as in the experiment. \nopagebreak

The onsite Coulomb repulsion $U$ itself does not suppress the extended
$s$-wave admixture induced at the boundary, but it can generate a finite
magnetization  $ m (\bf x)$ which then competes with this
superconducting state.
 If we look 
for a surface state with finite magnetization for the {\em regular FS}, 
we find indeed that already for $U=t$ its critical temperature is comparable
to that of the
$s+id$-wave state. For the {\em regular FS} parameters the weak spin
polarization (less than 3$\%$) is
ferrimagnetic and decays into the bulk on the scale of the
superconducting coherence 
length. It is accompanied by a TRSB spin-triplet $p$-wave admixture. 
The $p_y$-wave component is induced directly by the $ d $-wave state because the spin  
rotational symmetry is broken such that the total spin of the pair
is not a good quantum number anymore.  The $p_{x}$-wave component does not
appear, since it is odd with 
respect to specular reflection at the surface and is
suppressed by pair breaking. Note that
neither the magnetization nor the $p$-wave admixture lift the directional degeneracy.
Therefore the magnetic surface state does not generate charge or spin surface currents.
The split in the surface DOS by the magnetic state is shown in 
Fig.\ref{all_tp0}.  As a consequence of the lifted spin degeneracy of the 
ABS the surface DOS is different for spin-up (with respect to the 
magnetization axis) and spin-down electrons, a property which could be 
tested by spin-polarized quasiparticle tunneling. 

In Fig.\ref{udep} we show the $U$-dependence of the critical
temperatures for the two different surface instabilities. Already $U$ slightly
larger than $t$ yields a higher $T_c$ for the magnetic
surface state. In a narrow range of parameters the 
coexistence of $s+id$ and magnetic surface state is possible.

\begin{figure}  
\begin{center}
\includegraphics[width=.49\textwidth]{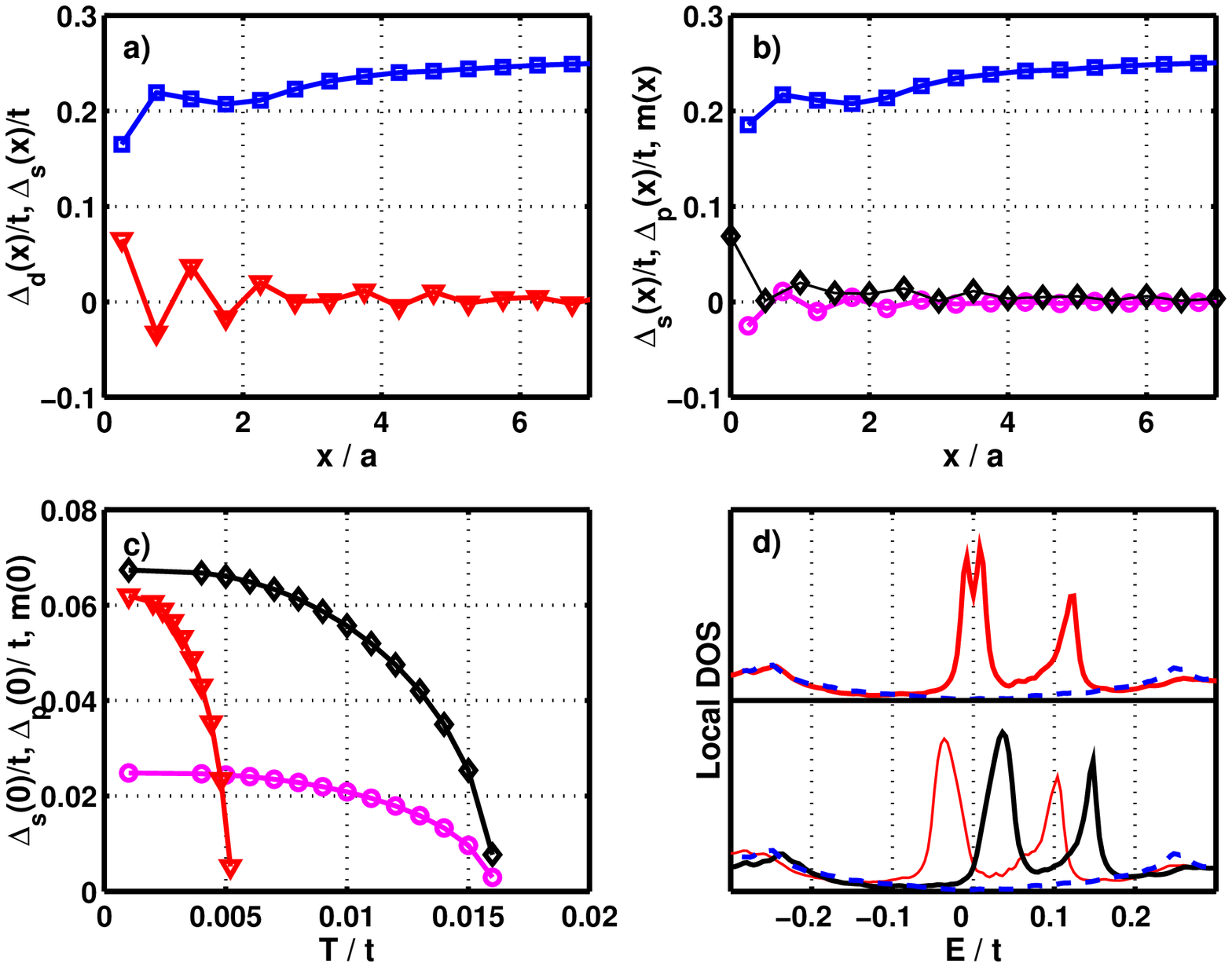} \end{center}
 \caption{Singular FS scenario (i.e. $t'=-0.3t$, $J=1.2t$, $\mu 
=-t$ and $T_c^d = 0.12 t$): a) $d$-wave gap (squares) and TRSB
$s$-wave admixture(triangles) at 
$T=0.001 t$ for $U=0$. b) $d$-wave gap, magnetization (diamonds)  and
TRSB $p$-wave admixture (circles)at 
$T=0.001 t$ for $U=t$. c) Temperature dependence of the $s$-wave gap function, $p$-wave
gap function  and 
magnetization on the first site. d) upper panel: Local DOS in
the $s+id$ state at the surface (solid line) and in the bulk (dashed
line); lower panel: Local DOS in 
the magnetic surface state at the surface (solid lines, spin-up and spin-down
DOS separate) and in the bulk (dashed line).}
\label{all_tp03}
\end{figure}

\begin{figure} \begin{center} 
\includegraphics[width=.48\textwidth]{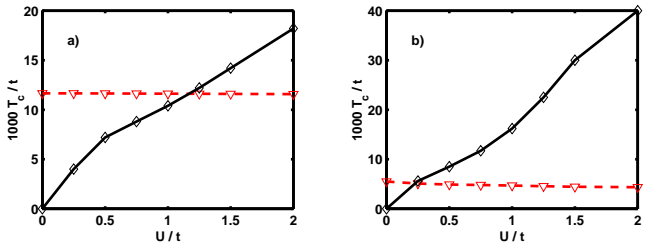}
\end{center} \caption{$U$-dependence of the critical temperature for the
$s+id$-wave (dashed line) and the magnetic surface state (solid line) for the
regular FS (a)) and the singular FS scenario (b)). }
\label{udep}
\end{figure}

Next we consider the case of the {\em singular FS} close to the VHS by choosing
$t'=-0.3t$ and $\mu = -t$. In order to obtain approximately the same value
for the $d$-wave gap magnitude as in the previous case, we take
$J=1.2t$. Due to the VHS a large 
part of the low-energy quasiparticles come from the $k$-space regions
around the $(\pi,0)$ and $(0, \pi)$ points (in crystal coordinates). Since the
wave function of the ABS is  $\propto \sin k_{Fx}x$ and $k_{Fx}
\approx \pi/a$ for most quasiparticles all quantities which live on
the 
bonds and involve products of wave functions on neighboring sites (at distance
$a/2$) oscillate like $\sin 2 k_{Fx}x$. This also holds for the
current carried by the bound states and the 
$s$-wave admixture at the surface. As  
a consequence, the surface currents carried 
by the ABS are nearly canceled and we do not find a transition
towards a pure $d$-wave SSC state down to very
low temperatures.  This is in
contrast to the results for the regular FS, that resemble those from
quasi-classical theory which is insensitive to
effects on such a microscopic length scale. Even if we admit a finite $s$-wave
coupling the critical temperature for the $s$ admixture
is drastically reduced compared with the previous case ($T_c^s \approx 
0.005
t$) (Fig.\ref{all_tp03}). The main reason for
the small $T_c^s$ is that the FS lies close to the 
node lines of the extended $s$-wave gap.
Due to the smaller $s$-wave admixture the split in
the ZEP for the $s+id$-wave state is rather weak (see
Fig.\ref{all_tp03} d)). 
Our results are in in qualitative agreement
with the results of Tanuma et al.\cite{tanuma} who use a $t-J$
model in Gutzwiller approximation at comparable band filling.  

For finite Coulomb repulsion $U=t$ we again find a magnetic surface state.
The VHS enhance correlations with the wavevector close to $ (\pi , \pi)
$ so that the magnetization
resembles a spin density wave with period $a$ decaying towards the
bulk region (see Fig.\ref{all_tp03} b)).
The induced $p_y$ gap component also exhibits $2k_F$ oscillations. We
obtain a sizable split in the surface DOS, the spin-resolved density of states
is shown in the lower panel of Fig.\ref{all_tp03}. We find also additional bound states
at higher positive energies. This is apparently an effect of the FS, the
vicinity to the VHS is responsible for the strong electron-hole
asymmetry. However these bound states exist in the entire $d$-wave phase and
are therefore not related to the low
temperature surface phase transitions. We would like to remark here: (1) the magnetization approaches the  
ideal staggered magnetization with period $a$ when we choose $t'=0$
and, additionally, stay close
to half filling; (2) the chosen values $U=t$ and $t'=-0.3t$ are
insufficient to establish a Neel state in the bulk in the absence of
superconductivity. However, the rearrangement 
of the ABS provides a mechanism to stabilize the 
magnetic surface state.
 
For the {\em singular FS} the critical temperature $T_c^m$ for the
magnetic surface state easily exceeds $T_c^s$ for the $s+id$-wave state (see
Fig.\ref{all_tp03} c) and Fig.\ref{udep}), so
that our results suggest that the magnetic surface state is the most stable
state in the considered frame of possibilities. However  an external magnetic
field creating a Doppler shift for the quasiparticles and therefore again
lifting the charge degeneracy of the ABS would support the $s+id$-wave state
in the competition with the seemingly quite robust magnetic surface
state. Note that the charge coupling corresponds to a considerably
higher energy scale than the Zeeman coupling which is negligible in
this case. This could induce a transition between these two
surface states as the external field is increased. We also refer the reader to 
a recent preprint by Hu and Yan\cite{huyan}, who  discuss possible giant magnetic
moments due to the split surface states.  

In summary, we have considered different mechanisms to explain the
observed low temperature splitting of the ZEP at
[110] surfaces of $d$-wave superconductors. On the one hand, we find
that a TRSB superconducting state leads to this effect
which is induced by the Doppler shift of a spontaneous surface current 
\cite{honunpub} or by the local admixture of an $s$-wave component ($
s + i d $) \cite{fogelstrom,yip}. On the other hand, electron correlation
effects lead to a magnetic instability related to the antiferromagnetic
state. Naturally, the latter is more stable in the underdoped regime
represented in our case by the model with a singular FS. The former
has a better chance to be realized in the overdoped region (regular
FS) where the antiferromagnetic spin fluctuations are sufficiently
reduced. The experimental distinction between the two states is possible
by spin-polarized tunneling as the magnetic state leads to a splitting 
of the surface DOS for up and down spin.

We are grateful to T.M. Rice, K. Kuboki, D. Agterberg and A. Fauchere
for many helpful discussions. We would like to thank for financial
support by the Swiss Nationalfonds and the Ministry of Education,
Science and Culture of Japan.


\vskip -.6 cm
\begin{references} 
\bibitem{josephson} D.J. van Harlingen, Rev.Mod.Phys. {\bf 67}, 515 (1995), and
references therein.
\bibitem{hu} C.R.Hu, Phys.Rev.Lett. {\bf 72}, 1526 (1995); S.Kashiwaya,
Y.Tanaka, M.Koyanagi, and K.Kajimura, Phys.Rev.B {\bf 53}, 2667 (1996). 
\bibitem{greene} L.H. Greene, M.Covington, M.Aprili, and E.Paraoanu, Solid State
Comm. {\bf 107}, 649 (1998); L.Alff et al., Phys.Rev.B {\bf 55}, R14757 (1997).
\bibitem{walter} H.Walter, W.Prusseit, R.Semerad, H.Kinder, W.Assmann,
H.Huber, H.Burkhardt, D.Rainer, and J.A.Sauls, Phys.Rev.Lett. {\bf 80}, 3598 (1998).
\bibitem{covington} M.Covington, M.Aprili, E.Paraonu, L.H.Greene, F.Xu,
J.Zhu, and C.A. Mirkin, Phys.Rev.Lett. {\bf 79}, 277 (1997).
\bibitem{fogelstrom} M.Fogelstr\"om , D.Rainer, and J.A.Sauls,
Phys.Rev.Lett. {\bf 79}, 281 (1997).
\bibitem{trsbreview} For a review see M. Sigrist,
Prog. Theor. Phys. {\bf 99}, 899 (1998).
\bibitem{yip} M. Fogelstrom and S.K. Yip, cond-mat/9803031.  
\bibitem{honunpub} S.Higashitani,
J.Phys.Soc.Jap. {\bf 66}, 2556 (1997); C.Honerkamp and M.Sigrist,
unpublished.
\bibitem{waka} M. Fujita, K. Wakabayashi, K. Nakada and K. Kusakabe, 
J. Phys. Soc. Jpn. {\bf 65}, 565 (1996).
\bibitem{tanuma} Y.Tanuma, Y.Tanaka, M.Ogata, and S.Kashiwaya,
J.Phys.Soc.Jap. {\bf 67}, 1118 (1998); see also J.X.Zhu, B.Friedman and
C.S.Ting, Phys.Rev.B {\bf 59}, 3352 (1999).
\bibitem{huyan} C.R.Hu and X.Z.Yan, cond-mat/9901334.
\end{references}
\end{document}